DEFINING THE PRINCIPLES FOR THE MOTION OF BLOOD THROUGH ARTERIES

*E 855 -- Principia pro motu sanguinis per arterias determinando*

Presented to the St. Petersburg Academy on December 21, 1775
Originally published (in an incomplete form) in Opera Postuma 2, 1862, pp. 814 - 823
Also Published (with the recovered missing parts) in Opera Omnia: Series 2, Volume 16, 1979, pp. 178 - 196

Translated and Annotated
by
Sylvio R. Bistafa[*]
December 2017

Foreword

As originally published in *Opera Postuma* 2, 1862, pp. 814-823, E 855 *Principia pro motu sanguinis per arterias determinando* of 1775 is an incomplete manuscript, whereas the complete version was published much later, in 1979, in *Opera Omnia*: Series 2, Volume 16, pp. 178 – 196 (see the Preface for more details about the missing parts). Truesdell[1] traces the mishaps of this publication, since its submittal to the first contest of Dijon Academy in 1742. The subject of the contest was "To determine the difference in velocities between a liquid that circulates through elastic and rigid tubes" (*Déterminer la différence des vitesses d'un liquide qui passe par des tuyaux inflexibles et de celui qui passe par des tuyaux élastiques*), whereas E 855's title, we suppose, would be more appealing to the context. In fact, as we shall see, although in the introduction to the problem (paragraphs 1 – 8), Euler focuses on the discussion on blood flow through arteries, to such an extent as to even propose *adhoc* models for the behavior of their elastic cross-sections, surprisingly, and also somewhat frustratingly, the main body of the manuscript (paragraphs 9 – 34) is devoted to the modeling of flows through rigid tubes, driven by a piston pump. It is only in the remaining sections of the manuscript (paragraphs 35 – 43) that Euler investigates the formulas for the motion of fluids in elastic tubes, ending the work with the very pessimistic philosophical statement that "[…] since there appears to be no way in which this can be completely resolved and the investigation can be considered to transcend human powers, the work will end here."

As Truesdell points out "[…] It is interesting that Euler's *first* paper on hydraulics should concern this extremely difficult subject […] since (E 855) employs concepts. Euler's paper was not to develop until 1755". In fact, what is considered to be Euler's first hydraulics paper appeared in printed version only in 1754[2].

Nonetheless, as Cerny and Walawender[3] point out "[…] This work [on blood flow] may be the first mathematical treatment of circulatory physiology and haemodynamics. Euler perhaps could properly be called the "Father of Haemodynamics". In this publication, Cerny and Walawender essentially translated the incomplete form of E 855, not on an equal footing with the original material, omitting most of the verbose parts, and adding comments and interpretations from a more modern point of view. Nonetheless, all the mathematical developments and the most relevant of Euler's assumptions and justifications were included. The authors succeeded in providing a very objective translation, covering the essential parts of Euler's manuscript on blood flow.

---

[*] Corresponding address: sbistafa@usp.br

[1] C. A. Truesdell, *Leonhardi Euleri, Commentationes Mechanica ad Theoriam Corporum Fluidorum Pertinentes, Volumen Posterius Set. Secunda XIII*, (Lausannae, Orell Pussli Turici, 1955), p. LXXVII.

[2] E206 -- *Sur le mouvement de l'eau par des tuyaux de conduite.* Originally published in *Mémoires de l'académie des sciences de Berlin* 8, 1754, pp. 111-148. Also in *Opera Omnia*: Series 2, Volume 15, pp. 219 – 250. According to C. G. J. Jacobi, a treatise with this title was presented to the Berlin Academy on October 23, 1749. For discussions on Euler's participation in practical hydraulic projects, see: M. Eckert. Euler and the Fountains of Sanssouci. Arch. Hist. Exact Sci. 56: 451–468, 2002, and S. R. Bistafa. Euler's friction of fluids theory and the estimation of fountain jet heights. Eur. Phys. J. H 40: 375–384, 2015.

[3] L. C. Cerny and W. P. Walawender. Leonhardi Euleri's *Principia pro motu sanguinis per arterias determinando*. J. Biol. Phys. 2: 41-56, 1974.

This translation differs from the aforementioned one, not only because it is based on the complete form of the manuscript, but also because, it is an integral translation, reflecting more faithfully the original Latin form of the manuscript in English. Annotations were also added where appropriate.

E 855 – Preface[4]

*Defining the principles for the motion of blood through arteries*
Posthumous works 2, 1862, p. 814 – 823 (p. 178-196)

The text of this work, presented in 1775 at the Academy of Petersburg, was only partially reproduced in the posthumous Works; since the editors only had in 1859, for the preparation of these posthumous volumes, an incomplete manuscript, lacking paragraphs 1 to 14; after that time, these were fortunately found in the archives of the Academy of Sciences of the URSS in Leningrad, which we have to acknowledge for their precious collaboration; these now appear in the present edition, for the first time in a printed form (the lack of § 6 is not due to an omission, but rather for an error of numeration).

The chosen model, led Euler to study the motion of a liquid in a tube with elastic walls. It is now known that he had already dealt with this problem, much earlier than 1775; yet in 1742, he had in fact sent to Dijon, in the occasion of a competition lunched by the Academy of this town, a submission that seems to have not reached its destination and that should be considered to be lost[5].

In regard to the chosen model (§ 1 to 8), Euler established the equations that satisfy the three functions $v$ (velocity), $p$ (pressure), $s$ (tube cross sections) of two variables $z$ (abscissa) and $t$ (time); the considerations then became classic, giving him the equation of continuity

$$\frac{dv}{dt} + v\frac{ds}{dz} + s\frac{dv}{dz} = 0,$$

and the said equation of acceleration

$$2g\frac{dp}{dz} + v\frac{dv}{dz} + \frac{dv}{dt} = 0,$$

to which he adds a finite relation linking $p$ and $s$ (as to neglect the inertia of the walls); he then supposes the existence of a section $\Sigma$ corresponding to an infinite pressure, such that the choice, obviously, among an infinite possible relations; without others, he then suggests

$$p = \frac{cs}{\Sigma - s},$$

and

$$p = c\, log\frac{\Sigma}{\Sigma - s} \qquad\qquad (§ 9\ to\ 17).$$

---

[4] *Opera Omnia*: Series 2, Volume 16, pp. XIV – XV (in French).
[5] This first work is mentioned in a letter of 3.12.1741 from A. C. Clairaut to Euler, published in *Opera Omnia* IV A5, p. 139; with regard to this subject see also the remarks by C. Truesdell in the preface of *Opera Omnia* II XIII p. LXXVII – LXXVIII.

Before considering the general problem, he considers the case of a rigid tube (§ 18 to 34), first by integrating the equations, and then by considering the momentum theorem. Finally, upon considering the general case (§ 35 to 43), he is faced with obstacles such that he concludes, here as well, for the insufficiency of analytical tools at his disposal. One may find in C. Truesdell's preface, a brief historical account on the researches done afterwards on the same subject.

## Defining the Principles for the Motion of Blood Through Arteries
______________________________
Comment 855 Index of Eneström
Posthumous Works 2, p. 814-823[6]

1. We revive that investigation on the physiological hydrodynamical principles, when we put a pump in the place of the heart, from which the fluid is expelled with the aid of a piston, flowing out from a certain place in the pump, such that the fluid runs alternately into the pump, and for this to happen, a certain force is applied to the piston. Truly, the tube leading to the arteries be in such way connected to the pump, which by itself be pressed by a force of contraction, such that not to leave any cavity, until a vacuum is formed, and eventually it will be expanded by a force applied by the fluid in transit, which will be as great as the state of the pressure in the fluid. This tube is laid straight as long as necessary, since from Hydrodynamics, the curvature of tubes, through which the fluid moves forward, hardly provokes any alteration in the motion. Furthermore, it will be enough to consider a unique continuous stretched tube and to remove all the ramifications, since we know that the motion through the various branches scarcely disturbs the fluid motion through the trunk from which theirs breadths are extended. Finally, where the arteries are inserted into tonsils or communicate with veins, we shall suppose an orifice in our tube that imposes a certain resistance to the fluid outflow, which corresponds to the transit difficulty of blood in veins or tonsils. Thus, it is manifested that the motion through such tube is not much different from the motion that is imposed to the blood through arteries.

2. Hence, we conceive that, in alternations, a certain quantity of fluid flows into our pump, meanwhile the tube exit remains closed, which, therefore, corresponds to the flow during the expansion of the heart, called diastole. So, thereafter, follows the contraction of the heart, called systole, which generates the fluid flow out of the pump through the tube, propelled by the action of the piston. Whence, the force that should be applied to the piston will be investigated, for the occurrence of the blood motion. Consequently, the true force of the heart will become known, which, on most occasions, is incorrectly attributed to the Physiologies. It is known that with the alternations of this pump, likewise, the motion of the fluid through our tube is closely submitted to this same alternation, such that it is alternatively spreaded out to the fluid running through, and collapses again, to be repressed by the reciprocal motion of the arteries pulse.

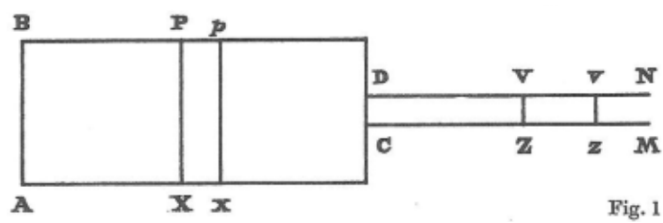

Fig. 1

---
[6] See the preface, p. XIV

3. Be then $ABMN$ our tube, connected with the pump in $AB$, the extremity $MN$ is truly an orifice through which the fluid is expelled; and since in the unique and particular point $Z$, its amplitude $ZV$ is variable, which in these circumstances will depend on the force applied by the pressure, which the fluid flows. It is necessary before anything else to define the maximum local amplitude of the tube in $Z$, that supports a maximum force, almost infinite in magnitude, which, therefore, in the various locations it can be greater or lesser, such that the amplitude of the arteries is increased or reduced, being strongest in the ramifications that derived in branches from the main cavity than in the cavity of the trunk. As a consequence, we remove these branches, being then convenient to allow a maximum amplitude to our continuous tube, as great as it is away from the heart. Then, from any point from which the distance to the pump we call $AZ = z$, we fix the maximum amplitude $= \Sigma$, such that the function should depend only on the variable $z$. In fact, this quantity will have two dimensions, since it expresses the amplitude which the tube in $Z$ can expand due an almost infinite force. Hence, if this amplitude is allowed to continuously increase in the direction of $MN$, it will be assigned at the beginning an amplitude $AB = aa$, so that it will be possible to write[7] $\Sigma = aa(1 + \alpha z^n)$, such that $\alpha$ is the fraction $\frac{1}{f^n}$ for maintaining the homogeneity. And yet, let us call the total longitude of the tube $AM = l$, and then the maximum amplitude of the orifice at the end will be $= aa\left(1 + \frac{l^n}{f^n}\right)$, whence from the real conditions of the arteries, it will be possible to conveniently assign the constant $f$ and the exponent $n$.

4. On the other hand, the amplitude $\Sigma$ at the location $Z$, will eventually be found, when the pressure of the fluid there is almost infinite, but it is never supposed to occur. Whence, it is clear that the pressure anywhere, which the flowing fluid transmits, will induce by itself and amplitude smaller than $\Sigma$, until it would become null when the pressure ceases; to increase towards $\Sigma$, once the pressure is continuously augmented.

5. Let us now find to which amplitude this tube is expanded in point $Z$, for a given pressure of the flowing fluid. Let us, then, assign the pressure of the fluid in this location to be as great as to sustain a body in such fluid submerged at depth $= p$, since this is the form in which the state of pressure is always defined in Hydrodynamics; then be $s$ the amplitude to which the tube in $Z$ is extended under such force, and of course, this letter designates a quantity in two dimensions, such as $\Sigma$. Then, the relation to be established between these amplitudes $\Sigma$ and $s$ will be such that being $p = 0$, $s = 0$ as well, and by putting $p = \infty$, gives $s = \Sigma$, which can be satisfied by an infinite number of conditions, among which the simplest is given by the formula $s = \frac{\Sigma p}{c+p}$, where $c$ denotes a constant quantity that depends on the degree of elasticity of our tube, such that if the elasticity would vanish, would be $c = 0$ as well, and we would have $s = \Sigma$, always; and therefore, there will be also the formula $p = \frac{cs}{\Sigma - s}$. Likewise, it is possible to create other innumerable formulas, which, in the same way, would satisfy the same conditions, from which it will be chosen in the following, the one which reveals itself the most suited to the calculation.

7.[8] With these considerations, let us establish now all the basic principles that should be brought forward in the determination of the fluid motion through the tube. Since indeed it was established that $AZ = z$, as the maximum amplitude of the tube in the $Z$ location $= \Sigma$, which is also expected to be a function of $z$ itself. Thus, once the time $t$ has elapsed, after the piston action had been applied, which we always express in seconds, being $v$ the velocity which the fluid flows through the tube in this location, at this moment. Of course, this velocity $v$ per second; nonetheless, it should be remembered that the letter $p$ is the height measuring the state of pressure of the fluid; and it is manifested that these two quantities $v$ and $p$ are, in fact, functions of two variables, obviously, the distance $AZ = z$, and the time $t$. Whence, furthermore, all the operations in order to establish these functions of two variables will now require to be established. Moreover, the true expression for $p$ will furnish the true amplitude $s$ which the tube in $Z$ will have at this time, which can be obtained with the aid of the formula such as $s = \frac{\Sigma p}{c+p}$, or another equivalent one. Thus, it follows that the letter $s$ will be a function of two variables $z$ and $p$.

---

[7] Uncertain reading.
[8] § 6 was omitted.

8. Thus, all our approach here allows us to be able to evaluate both functions $v$ and $p$, for any interval $AZ = z$ and for any time $= t$, which, in fact, will require the investigation of the two necessary equations, one which should be sought from the continuity of the fluid flowing through the tube, and the other, in fact, from the acceleration originated from the forces pressing a single fluid element, wherefore, they will be extracted from two of our principles, one of the continuity, and the other, in fact, from the acceleration, from which the true whole motion of the fluid through such tubes will be required to determine.

## I. On the Principle of Continuity

9. Then, let us consider the fluid element occupying the tube space $ZVvz$ at a certain instant of time, where $ZV$ refers to the actual amplitude $s$ to which the tube in $Z$ is now expanded, such that be established that the element $Zz = dz$, and that the mass of the fluid in this small space $= sdz$; also the amplitude in a near location will be $zv = s + ds$, where the variability of the element $ds$ should be assumed to depend on $z$ itself, which, as such, is represented, as usual, as $dz\left(\frac{ds}{dz}\right)$, in this way, it will be $zv = s + dz\left(\frac{ds}{dz}\right)$. In a similar way, since the velocity of the fluid in $ZV$ at this instant is $= v$, the velocity in $zv$ at this same instant is $= v + dz\left(\frac{dv}{dz}\right)$.

10. Now, let us assume that after the infinitesimal time interval $dt$ has elapsed, the mass of the fluid $ZVvz$ is brought to the position $Z'V'v'z'$, and since it necessarily follows that the volume is now $= sdz$, then, our equation will be generated from this condition. Therefore, once considered that in $ZV$, the velocity is $= v$, and that in the infinitesimal time interval $dt$, it is transported to $Z'V'$, covering the distance $ZZ'$, then it is necessary that $ZZ' = vdt$. In a similar way, since the velocity attached to $zv$ is $v + dz\left(\frac{dv}{dz}\right)$, and from there it is transported to $z'v'$, the traversed distance $zz'$ will be

$$zz' = vdt + dtdz\left(\frac{dv}{dz}\right),$$

from which the distance $Zz'$ will be

$$Zz' = dz + vdt + dtdz\left(\frac{dv}{dz}\right).$$

Hence, once the distance $ZZ' = vdt$ is subtracted from $Zz'$, it will result in the infinitesimal distance $Z'z'$

$$Z'z' = dz + dtdz\left(\frac{dv}{dz}\right).$$

11. Let us now also look for the amplitude of the tube $Z'V'$, which will originate from the original $ZV = s$, as the increment of the variable $z$ assumes a value $= vdt$, simultaneously with the time increment $dt$ applied at time $t$, whence both variabilities give $ds = dz\left(\frac{ds}{dz}\right) + dt\left(\frac{ds}{dt}\right)$; and by putting $vdt$ in place of $dz$, then the unknown will be

$$Z'V' = s + vdt\left(\frac{ds}{dz}\right) + dt\left(\frac{ds}{dt}\right),$$

whose value once multiplied by the distance $Z'z'$, will result in the volume given by the equation

$$Z'V'Z'z' = sdz + dzdt\left[v\left(\frac{ds}{dz}\right) + s\left(\frac{dv}{dz}\right) + \left(\frac{ds}{dt}\right)\right],$$

and since it should be equal to the preceding volume $sdz$, we finally find the equation

$$0 = v\left(\frac{ds}{dz}\right) + s\left(\frac{dv}{dz}\right) + \left(\frac{ds}{dt}\right),$$

which can be written as

$$\left(\frac{ds}{dt}\right) + \frac{d(vs)}{dz} = 0,$$

and this is the equation that our principle of continuity furnishes.

## II. On the Principle of Acceleration

12. Since $v$ is the velocity applied to $ZV$ at time $t$, after the infinitesimal time $dt$ it reaches the position $Z'V'$, travelling the distance $ZZ' = vdt$, and from the variability of both variables $z$ and $t$, this will give the velocity applied to $Z'V'$, once $dz$ is written as $vdt$, whence that velocity will be

$$= v + vdt\left(\frac{dv}{dz}\right) + dt\left(\frac{dv}{dt}\right).$$

Hence, this increment of the velocity, once divided by $dt$, will give the acceleration of the fluid applied to $ZV$, and this will be

$$= v\left(\frac{dv}{dz}\right) + \left(\frac{dv}{dt}\right).$$

13. As a consequence, this acceleration should be produced by the forces with which the element of the fluid $ZVzv$ is driven at this instant of time. In fact, the anterior face $ZV$ of this element endures the pressure due to the height $p$, whereas the posterior face $zv$ will endure the pressure due to the height $p + dp$, with the differential $dp$ taken only from the variability of $z$ itself, which then is given by $dz\left(\frac{dp}{dz}\right)$, such that the total pressure that the face $zv$ endures, be due to the height $p + dz\left(\frac{dp}{dz}\right)$; the water, therefore, is pressed backwards by the pressure due to the height $dz\left(\frac{dp}{dz}\right)$, which is applied to the amplitude $s$ itself, and, then, the basis in which it acts, will be submitted to the driving force $sdz\left(\frac{dp}{dz}\right)$, expressed, of course, per unit of volume, whose weight is equated to the driving force; so that if this force is divided by the mass $sdz$ in which it acts, will result in an accelerating backward driving force $= \left(\frac{dp}{dz}\right)$, which according to the principle of Mechanics, that it behooves the acceleration to be divided by $2g$, denoting $g$ the height that a weight falls in one second[9]. Consequently, the equation from the principle of acceleration will be

$$2g\left(\frac{dp}{dz}\right) = -v\left(\frac{dv}{dz}\right) - \left(\frac{dv}{dt}\right).$$

14. See then our both equations, which our principles of continuity and of acceleration have furnished, which, therefore, will hold as

$$I. \left(\frac{ds}{dt}\right) + \frac{d(vs)}{dz} = 0,$$

---

[9] TN: the height that a weight falls in on second is the acceleration of gravity divided by two; hence the actual acceleration of gravity is $2g$.

$$II. \; 2g\left(\frac{dp}{dz}\right) + v\left(\frac{dv}{dz}\right) + \left(\frac{dv}{dt}\right) = 0^{10},$$

and to which it is added the formula

$$s = \frac{\Sigma p}{c+p} \quad \text{or} \quad p = \frac{cs}{\Sigma - s},$$

and these should allow to deduce what is pertinent for determining the motion of the fluid through the tube.

15.[11] Now, since there is a relation between $p$ and $s$, it will be necessary, thence, to obtain the value for the expression $\left(\frac{dp}{dz}\right)$, which, when written as a function of $z$ alone, will result in

$$\frac{dp}{dz} = \frac{c}{(\Sigma - s)^2}\left[\Sigma\left(\frac{ds}{dz}\right) - \frac{sd\Sigma}{dz}\right],$$

and this can concisevely be presented as

$$\frac{dp}{dz} = \frac{c\Sigma^2}{(\Sigma - s)^2} \frac{d\left(\frac{s}{\Sigma}\right)}{dz}.$$

And thus, the previous equation assumes the form

$$\frac{2gc\Sigma^2}{(\Sigma - s)^2} \frac{d\left(\frac{s}{\Sigma}\right)}{dz} + v\left(\frac{dv}{dz}\right) + \left(\frac{dv}{dt}\right) = 0,$$

such that now there are only two variables $s$ and $v$ that should be determined from the principle variables $z$ and $t$.

16. Thus, we have unfolded the two main equations, which we write as

$$I. \; v\left(\frac{ds}{dz}\right) + s\left(\frac{dv}{dz}\right) + \left(\frac{ds}{dt}\right) = 0 \quad \text{and} \quad II. \left(\frac{dv}{dt}\right) + v\left(\frac{dv}{dz}\right) + \frac{2gc\Sigma^2}{(\Sigma - s)^2} \frac{d\left(\frac{s}{\Sigma}\right)}{dz} = 0.$$

Now, a posterior introduction of $s$, after the introduction of $v$ we get the following equation

$$s\left(\frac{dv}{dt}\right) - v^2\left(\frac{ds}{dz}\right) - v\left(\frac{ds}{dt}\right) + \frac{2gcs\Sigma^2}{(\Sigma - s)^2} \frac{d\left(\frac{s}{\Sigma}\right)}{dz} = 0,$$

which will be employed in a posterior section.

17. But, if in place of the hypothesis that we have $s = \frac{\Sigma p}{c+p}$, we desire to use this $s = \Sigma\left(1 - e^{-\frac{p}{c}}\right)$, then the same phenomena should approximately be reproduced, whence it would give

$$p = c \log \frac{\Sigma}{\Sigma - s}, \qquad \text{and hence} \qquad \left(\frac{dp}{dz}\right) = \frac{c\Sigma}{\Sigma - s} \frac{d\left(\frac{s}{\Sigma}\right)}{dz},$$

which is a considerable simpler expression than the one that was given by

$$\frac{c\Sigma^2}{(\Sigma - s)^2} \frac{d\left(\frac{s}{\Sigma}\right)}{dz},$$

and thus, it can be safely adopted.

*Developing the case where the tube is assumed to be rigid*

---

[10] TN: the equation that resulted from the Principle of Acceleration, is recognized as what is now called Euler's equation for unsteady, one-dimensional flows.

[11] Here begins part of those commentaries introduced in <<posthumous Works>>. C.B.

18. Since the solution of these equations is unknown, it will be convenient to begin with a well known case, that of a tube through which the fluid is propelled is assumed to be rigid, and thus, its amplitude $s$ is the maximum amplitude $\Sigma$; and for this reason, a function of the variable $z$ only. In this case, we have that $\left(\frac{ds}{dt}\right) = 0$, whence the two equations determining the motion will be

$$I.\ \frac{d(vs)}{dz} = 0. \qquad II.\ 2g\left(\frac{dp}{dz}\right) = -\left(\frac{dv}{dt}\right) - v\left(\frac{dv}{dz}\right).$$

19. The first of these equations, in which $\frac{d(vs)}{dz}$ is considered to be a function of $z$ only, as long as the time $t$ remains constant, upon integration gives $vs = T$, denoting $T$ a function of $t$ itself, which should be the same in any location of the tube, and if in the location where the piston propels the fluid is $= b$, and the velocity which the piston moves $= V$, it is manifested that, if it is transported to the location of the undefined point $Z$, then the amplitude $s$ is transformed into $b$, and particularly, the velocity $v$ into $V$, where $V$ depends on the time $t$ only, and thus $T$ will be $= bV$, such that the integral of the first equation gives us $sv = bV$, and hence $v = \frac{bV}{s}$; and, obviously, in the particular $Z$ location of the tube, the velocity of the fluid will be inversely proportional to the amplitude $s$ of the tube, which is a direct result that the theory of the motion of fluids though tubes postulates.

20. Thus, from the first equation we have found $v = \frac{bV}{s}$, where $V$ is a function of the time $t$ only, $s$, also a function of the quantity $z$ only, and then, let us substitute this value in the other equation, and from

$$\left(\frac{dv}{dt}\right) = \frac{b}{s}\frac{dV}{dt} \qquad \text{and} \qquad \left(\frac{dv}{dz}\right) = -bV\frac{ds}{s^2 dz},$$

will result in

$$2g\left(\frac{dp}{dz}\right) = -\frac{b}{s}\frac{dV}{dt} + b^2V^2\frac{ds}{s^3 dz},$$

which is appropriate to find the pressure $p$, whose differential $dp$, is related to the unique variable $z$, meanwhile the time $t$ in this equation remains constant; and from the quantities that depend on time only, meaning $V$ and $\frac{dV}{dt}$, whence once multiplied by $dz$, will then transform the equation into

$$2g dp = -\frac{b dV}{dt}\frac{dz}{s} + b^2V^2\frac{ds}{s^3},$$

which once integrated will give

$$2gp = -\frac{b dV}{dt}\int \frac{dz}{s} - \frac{1}{2}b^2V^2\frac{1}{s^2} + T,$$

where $T$ designates that function of time, which will be defined according to the specific investigation, and to its determination, we transport the point $Z$ to the location where the piston operates; which then gives $s = b$, and also taken the integral $\int \frac{dz}{s}$ to vanish in this location; then the pressure $p$ becomes equal to the pressure which the piston acts in the fluid, and whose force is supposed $= P$, and the resulting equation in the location of the piston will be $2gP = -\frac{1}{2}V^2 + T$, whence the result is $T = 2gP + \frac{1}{2}V^2$, which when substituted into our other equation will give

$$2gp = -\frac{b dV}{dt}\int \frac{dz}{s} - \frac{1}{2}b^2V^2\frac{1}{s^2} + 2gP + \frac{1}{2}V^2,$$

or

$$2gp = 2gP + \frac{1}{2}V^2\left(1 - \frac{b^2}{s^2}\right) - \frac{b dV}{dt}\int \frac{dz}{s}.$$

21. (Fig. 1) Let us now consider the pump $ABCD$, which at the beginning will be refilled with fluid, meanwhile the annexed tube $CDNM$ had been totally in vacuum, and be established that the amplitude of the pump is everywhere $= b$. Hence, after the time $= t$ has elapsed, and the action of the piston had protruded the fluid such that it now occupies the space $XPMN$, and the mass of the fluid, which at the beginning had occupied the space $ABPX$, fills now the tube $CDMN$. Let us then put the distance $AX = X$, such that the space $ABPX$ be $bX$. Hence, the piston that moves in section $PX$, exerts a pressure due to the height $P$, such that its total force is equivalent to the weight of the total mass of the fluid whose volume is $= bP$. Thus, in this way, the letter $P$ will denote the pressure itself, which the fluid supports in $XP$, as we have already assumed. Hence, since $V$ denotes the velocity of the fluid in $PX$, which in the infinitesimal time $dt$ is promoted through the infinitesimal distance $dX$, then, $dX = Vdt$, and thus $X = \int Vdt$. It is also evident to consider that the quantities $V$, $P$ and $X$ should also be functions of time $t$ only. Under these conditions, the volume $CDMN$, which now the fluid occupies in the tube will be $= bX = b\int Vdt$, evidently taken the integral $\int Vdt$ such that it vanishes for $t = 0$. Be further established that the total length of the pump $AC = a$, which everywhere has the same amplitude $b$, the total mass of the fluid, in which the piston acts will be $= ab$, where it is observed that $b$ is a quantity of two dimensions.

22. Now, let us adapt the more simple equation that was found to the case where we put the inlet of the tube connected to the pump in $CD$, and let us call the interval $CZ = z$, such that in $Z$ be the amplitude $ZV = s$ and the pressure $= p$, being the velocity that we already found $v = \frac{bV}{s}$. Be now $Z$ the value of the integral $\int \frac{bdz}{s}$, put to vanish at $z = 0$. Let the point $Z$[12] where we had assumed that this integral is supposed to vanish be transferred to $X$, then it is evident that

$$\int \frac{bdz}{s} = Z + a - X = Z + a - \int Vdt. \tag{1}$$

Let us put this expression put in place of $\int \frac{bdz}{s}$ and our equation will be

$$2gp = 2gP + \frac{1}{2}V^2\left(1 - \frac{b^2}{s^2}\right) - \frac{dV}{dt}(Z + a - \int Vdt), \tag{2}$$

this equation is applied to the total mass of the fluid contained in the tube, from the undefined point $Z$ to the point $C$ and it will have to be advanced all the way towards $M$, such that the volume $CDMN$, which is $\int sdz$ with $z = CM$, equals the volume that escapes $bX = b\int Vdt$, such that this equation would become appropriate to investigate the total interval $CM$, which obviously cannot be generally performed, unless the relation between the amplitude $s$ and $z$ is known.

23. Let us put the interval $CM = \omega$, the amplitude in this location $MN = \sigma$, and pressure $= \pi$; furthermore, the actual value of the integral $\int \frac{bdz}{s} = Z$, and once put $z = \omega$, it will be transformed into $\Omega$, and thus, for the case which we know handle, the integral $\int sdz$, from $z = 0$ up to $z = \omega$ should then become $= ab$. In fact, these posteriors denominations can also be applied when the fluid has already filled up the tube and flows in the orifice $MN$, and in this case the quantities $\omega$, $\sigma$ and $\Omega$ will be constant. Here, however, we will consider that these are functions of time $t$; meanwhile the pressure $\pi$ in this location, if the fluid is here circumscribed or if it flows, can always be considered to be known, and in this circumstance can be inserted for the determination of the fluid motion, as long as another unknown remains unresolved.

24. Thus, let us thoroughly determine the motion of the fluid, extending the undefined point $Z$ all the way to $M$, and putting $z = \omega$, $Z = \Omega$, $s = \sigma$ and $p = \pi$, such that we will get this equation

---

[12] TN: for the resulting integral to make sense, the point in question should be $C$, not $Z$.

$$2g\pi = 2gP + \tfrac{1}{2}V^2\left(1-\tfrac{b^2}{\sigma^2}\right) - \tfrac{dV}{dt}(\Omega + a - \int Vdt), \qquad (4)$$

where now all the quantities are functions of the time $t$ only, among then, just the velocity $V$ is expected to be the unknown, whose value is necessary to define, where it is seen that the only difficult to remain, is the integral $\int Vdt$. It is then necessary, that we transform this equation into another form, which enables an easier handling, and that it can be most suitable achieved, if we introduce in the place of the time $t$, the interval $AX = X$ itself, when besides $X$ being a function of the time $t$ only; then, moreover, in face of $\int Vdt = X$, there will be $dt = \tfrac{dX}{V}$.

Then it follows the equation

$$2g\pi = 2gP + \tfrac{1}{2}V^2\left(1-\tfrac{b^2}{\sigma^2}\right) - \tfrac{VdV}{dX}(\Omega + a - X), \qquad (5)$$

which, if we put $V^2 = 2y$, transforms into

$$2g\pi = 2gP + y\left(1-\tfrac{b^2}{\sigma^2}\right) - \tfrac{VdV}{dX}(\Omega + a - X), \qquad (6)$$

whose solution presents no great difficulty, which for this matter, we can now consider that now the velocity $V$, to be a function of the known quantity $X$.

25. Once the last equation has been solved, the time $t$ can be easily introduced again, and since now the velocity $V$ can be written as a known function of time, the fluid pressure can be conveniently ascribed in any location of the tube. Since the last equation gives

$$\tfrac{dV}{dt} = \frac{2g(P-\pi)+\tfrac{1}{2}V^2\left(1-\tfrac{b^2}{\sigma^2}\right)}{\Omega+a-X}; \qquad (7)$$

whose value once substituted in the equation for the pressure will give the following expression

$$2gp = 2gP + \tfrac{1}{2}V^2\left(1-\tfrac{b^2}{s^2}\right) - \frac{2g(P-\pi)+\tfrac{1}{2}V^2\left(1-\tfrac{b^2}{\sigma^2}\right)}{\Omega+a-X}(Z+a-X), \qquad (8)$$

or else, the calculation will turn out more elegant as follows. Since we have these two equations

$$2gp = 2gP + \tfrac{1}{2}V^2\left(1-\tfrac{b^2}{s^2}\right) - \tfrac{dV}{dt}(Z+a-X), \qquad (9)$$

$$2g\pi = 2gP + \tfrac{1}{2}V^2\left(1-\tfrac{b^2}{\sigma^2}\right) - \tfrac{dV}{dt}(\Omega+a-X), \qquad (10)$$

the difference of these equations gives us

$$2g(p-\pi) = 2gP + \tfrac{1}{2}b^2V^2\left(\tfrac{1}{\sigma^2}-\tfrac{1}{s^2}\right) - \tfrac{dV}{dt}(Z-\Omega), \qquad (11)$$

from which it is possible to define the pressure $p$ of the fluid in any location $Z$ of the tube at any time; and all that is desired related to the motion will be known from this calculation.

*Application to a tube with the same amplitude everywhere*

26. Let us assume that the amplitude of the tube $CM$ is everywhere the same $= c$, such that $s = c$, and thus, $\int sdz = cz$, whence once made $z = \omega$ it will result in $c\omega = bX$; and once put $c = nb$, will be $\omega = \tfrac{X}{n}$. Then, furthermore, actually will be $\sigma = c = nb$; furthermore, on the other hand, it will be considered

$$Z = \int \tfrac{bdz}{c} = \tfrac{z}{n}, \text{ hence } \Omega = \tfrac{\omega}{n} = \tfrac{X}{n^2}.$$

Since, in fact, in the location $M$ where the fluid is delimited the pressure is null, will be $\pi = 0$, whence our equation[13], once written in this location will be

$$0 = 2gP + \tfrac{1}{2}V^2\left(1 - \tfrac{1}{n^2}\right) - \tfrac{VdV}{dX}\left(a + \tfrac{X}{n^2} - \mathrm{X}\right). \tag{12}$$

Thereafter, in fact, for any undefined location $Z$ the equation[14] gives

$$2gp = 2gP + \tfrac{1}{2}V^2\left(1 - \tfrac{1}{n^2}\right) - \tfrac{VdV}{dX}\left(a + \tfrac{z}{n} - \mathrm{X}\right). \tag{13}$$

27. Among these prior equations, once taken the one particularized for the boundary $M$[15], which is as such only dependent on the variable $X$, after the multiplication by $2dX$, would then be given by

$$0 = 4gPdX + \tfrac{1}{2}V^2\left(1 - \tfrac{1}{n^2}\right) - 2VdV\left(a + \tfrac{X}{n^2} - \mathrm{X}\right), \tag{14}$$

which upon integration results in

$$C = 4g\int PdX - aV^2\left(1 - \tfrac{1}{n^2}\right) + V^2\left(1 - \tfrac{1}{n^2}\right)X, \tag{15}$$

whence, we get

$$V^2 = \tfrac{C - 4g\int PdX}{\left(1 - \tfrac{1}{n^2}\right)X - a}, \tag{16}$$

now, if it is assumed that the force $P$ of the piston is a constant $= B$, and the motion had begun where $X = 0$, and the velocity $V = 0$, because $\int PdX = BX$ the constant will be $C = 0$, and, hence, it will be

$$V^2 = \tfrac{4gBX}{a - \left(1 - \tfrac{1}{n^2}\right)X}, \tag{17}$$

whence, when the piston is propelled throughout the pump or the distance $AC = a$, so that once put $X = a$, then will be $V^2 = 4n^2 gB$; and also will be $CM = \tfrac{a}{n}$, this is also the velocity that successively the fluid advances uniformly through the tube[16], until a new action of the piston advances the flowing fluid.

28. Moreover, withstanding the action of the piston, whose force we assume to be constant, is the pressure $p$, which can be easily determined in any location $Z$ of the tube by means of an equation presented earlier in paragraph 26. If the difference is taken from them[17], the resulting equation will be

$$2gp = \tfrac{VdV}{dX}\left(\tfrac{X}{n^2} - \tfrac{z}{n}\right). \tag{18}$$

Since it is

$$V^2 = \tfrac{4gBX}{a - \left(1 - \tfrac{1}{n^2}\right)X}, \tag{19}$$

upon differentiation

$$\tfrac{VdV}{dX} = \tfrac{2agB}{\left[a - \left(1 - \tfrac{1}{n^2}\right)X\right]^2}, \tag{20}$$

---

[13] TN: actually from Eq. 10, substituting $\tfrac{VdV}{dX}$ for $\tfrac{dV}{dt}$, and $\tfrac{X}{n^2}$ for $\Omega$.
[14] TN: actually from Eq. 9, substituting $\tfrac{VdV}{dX}$ for $\tfrac{dV}{dt}$, and $\tfrac{z}{n}$ for $Z$.
[15] TN: actually Eq. 12.
[16] TN: here should be the 'pump' not the 'tube', since $V$ refers to the velocity of the piston.
[17] TN: the difference between Eq. 13 and Eq. 12.

And once this value is substituted, the pressure will be given by

$$p = \frac{aB}{\left[a-\left(1-\frac{1}{n^2}\right)X\right]^2}\left(\frac{X}{n^2} - \frac{z}{n}\right). \tag{21}$$

These expressions can be rendered simpler, by putting $\frac{1}{n} = \lambda$, such that $b = \lambda c$, where the number $\lambda$ can always be considered greater than unit, and thus, in fact, we will have

$$V^2 = \frac{4gBX}{a+(\lambda^2-1)X}, \tag{22}$$

and

$$p = \frac{aB}{[a+(\lambda^2-1)X]^2}(\lambda^2 X - \lambda z). \tag{23}$$

Since the velocity in the undefined location $Z$ is in evidently $v = \lambda V$, whence it will happen that

$$v^2 = \lambda^2 V^2 = \frac{4\lambda^2 gBX}{a+(\lambda^2-1)X}. \tag{24}$$

29. But, if we now examine carefully all the process of these operations, we will find out that contributed much to the success, the elimination of the time $dt$, which in its place we wrote $\frac{dX}{V}$, after that, we naturally eliminated the velocity $v$ from the calculations, and all the work could be conducted such that the velocity of the piston $V$ as a function of the variable $X$ could be defined. Then, in fact, the pressure $p$ turned out to be a function of the two variables $X$ and $z$. Besides, it is here pointed out with all the due importance, that thereafter, from a unique equation, not only the velocity $V$, but the pressure $p$ as well, could be defined, which is given as a function involving two natural variables. From this, it is recognized that the investigation of the motion through elastic tubes should also be undertaken.

30. But, if also this treatment we had wished to undertake in a similar way, we will be faced with quite inextricable calculations; because the integration of the two principal formulas is not easily accomplished. However, here, for not little it brings us a paradox, so in that it consist of, whilst naturally containing two equations for the whole solution for the case of rigid tubes

$$1)\ vs = bV,$$

and

$$2)\ 2gp = 2gP + \frac{1}{2}V^2\left(1 - \frac{b^2}{s^2}\right) - \frac{VdV}{dX}(Z + a - X),$$

obtained from the integration of the principal equations, likewise, by another way beginning with a different method may give an indication, to what degree a more careful analysis will be worth to be developed, whose outcome could be successfully applied to elastic tubes.

*Another method for determining the motion through rigid tubes*

31. Maintaining all the notation adopted thus far, let us consider the undefined portion of the fluid contained in the space $XP$ and $ZV$, whose volume will be $b(a - X) + \int sdz$, meanwhile, of course, the integral $\int sdz$ is taken from the boundary $z = 0$. In infinitesimal time let this portion be now displaced to the location $xpzv$, arising the infinitesimal distance $Xx = dX$ and $Zz = dz$, which is traversed with velocity $V$ and $v$, such that $\frac{dX}{dz} = \frac{V}{v}$, and $dz = \frac{vdX}{V}$; and on account of the principle that the mass of this portion should be constant, the differential of the formula $b(a - X) + \int sdz$, assuming variable with $X$ as well as with $z$, should equate to zero, and if $\frac{vdX}{V}$ is written in the place of $dz$; and since $s$ is a function of $z$ only, and likewise the formula $\int sdz$ itself, the derivative will result in $-bdX + \frac{vsdX}{V} = 0$, and therefore $vs - bV = 0$ or $vs = bV$.

32. To find the other equation[18], let us consider the living force of the same portion of the fluid $XPVZ$, which is given by the multiplication of this single element by the square of the velocity with which it is moved, whence that living force will be $= b(a - X)V^2 + \int v^2 sdz$, and if the value just found is substituted in the place of $v$, will result in

$$b(a - X)V^2 + b^2 \int \frac{V^2 dz}{s}, \tag{25}$$

in this expression, $V$ is independent of $z$, and thus can be rewritten as

$$b(a - X)V^2 + b^2 V^2 \int \frac{dz}{s}. \tag{26}$$

In case this mass is moved forward to the next position, such that $dz = \frac{vdX}{V}$ or $dz = \frac{bdX}{s}$, the increment of the living force originated at this moment will be

$$2bVdV\left(a - X + \int \frac{bdz}{s}\right) + bV^2\left(-dX + \frac{b^2 dX}{s^2}\right)$$
$$= 2bVdV(a - X + Z) + bV^2 dX\left(-1 + \frac{b^2}{s^2}\right). \tag{27}$$

33. For shortness, let us call this increment of the living force $WdX$ and then

$$W = \frac{2bVdV}{dX}(a - X + Z) + bV^2\left(-1 + \frac{b^2}{s^2}\right), \tag{28}$$

such that the other equation which previously by integration we found to be given by

$$2gP - 2gp = \frac{VdV}{dX}(a - X + Z) - \frac{1}{2}V^2\left(1 - \frac{b^2}{s^2}\right), \tag{29}$$

introducing now into this equation the increment of the living force $WdX$, gives the following $2gP - 2gp = \frac{W}{2b}$, such that the increment of the living force itself is

$$WdX = 4g(P - p)dX, \tag{30}$$

which in an excellent way, agrees with the principle of motion applied to the generation of the living force, where $P$ represents the force propelling the fluid, and $p$ the force retro propelling it[19].

34. For a clearer appreciation of the matter, let us consider, in general, the mass $M$ moving with velocity $V$, and a driving force $\Pi$ applied in the same direction, and since the first principle of motion gives $MdV = 2g\Pi dt$, whence, if the infinitesimal displacement is put $= dX$ such that $dt = \frac{dX}{V}$, this equation will become $MVdV = 2g\Pi dX$ or $2MVdV = 4g\Pi dX$, where $2MVdV$ is manifestly the increment of the living force $MV^2$, which, consequently, is always equated to the expression $4g\Pi dX$. Also, in our case, we have two propelling forces, one is $= bP$, which is actually applied through the infinitesimal distance $dX$, whence it generates the living force $= 4gbPdX$, while the other opposite force $= ps$, which once applied through the infinitesimal distance $dz$, generates the living force $4gspdz$, which, since $dz = \frac{bdX}{s}$, will be given by $4gbPdX$ that after subtraction leaves the true increment of the living force $4gbdX(P-p)$, which certainly should be equal to $WdX$ itself, as we straightforwardly found above. And thus, likewise, the equation which we found above by integration, we had obtained it now by differentiation. Likewise, it will be also appropriate to consider the case of elastic tubes.

---

[18] TN: Equation 2).

[19] TN: if it is considered that $W$ is a force, then, $= m\frac{dv}{dt}$, and the work done by this force $WdX = mvdv$, which upon integration, on the assumption that $W$ is constant, and the mass $m$ is accelerated from rest, gives $WX = m\frac{v^2}{2}$. The left-hand side is the work done by the force $W$, whereas the right side is the acquired kinetic energy (akin to the living force $mv^2$) during the displacement $X$. Then, it is seen that, in fact, $WdX$ is the work done by the force $W$ on $m$, during the infinitesimal displacement $dX$.

*Investigation of Formulas for the Motion of Fluids through Elastic Tubes*

35. Here, therefore, be established right onwards the same rationality, with the only difference that the amplitude $s$ is a function of two variables $X$ and $z$, such that its differential is

$$ds = dX\left(\frac{ds}{dX}\right) + dz\left(\frac{ds}{dz}\right); \tag{31}$$

then, let us consider the quantity of fluid contained in the undefined space $XPZV$, which, as before, is $b(a-X) + \int sdz$, whose increment is then obtained out of both variables $X$ and $z$, and given by $-bdX + sdz + dX\int dz\left(\frac{ds}{dX}\right)$, which, of course, expresses the difference in volume between the space $xpzv$ and the space $XPZV$.

36. Since, in fact, the velocity in $XP$ is $= V$, in $ZV$, truly, $= v$, while the stratum $ZV$[20] advances through the space $Xx$, the other stratum $ZV$ will be extended to the corresponding space $Zz = dz$, resulting in $\frac{dX}{dz} = \frac{V}{v}$, therefore, $dz = \frac{vdX}{V}$, then, if in the above expression in the place of $dz$ we write this value $\frac{vdX}{V}$, it should be equal to zero, whence the following result arises $-b + \frac{sv}{V} + dX\int dz\left(\frac{ds}{dX}\right)[=0]$, and by isolating the velocity $v$ in the left-hand side of this equation, will result in

$$\frac{bV}{s} - \frac{V}{s}\int dz\left(\frac{ds}{dX}\right). \tag{32}$$

For shortness, let us put $\int dz\left(\frac{ds}{dX}\right) = S$, to give $v = \frac{V(b-S)}{s}$; giving, then, the infinitesimal space

$$Zz = \frac{(b-S)dX}{s}. \tag{33}$$

37. We will also contemplate now the living force of the fluid contained in the space $XPZV$, which, as before, it will be given by $bV^2(a-X) + \int v^2 sdz$, and once $V$, which is a function of $X$ itself, is put in the place of $v^2$, it will result in this form

$$V^2\left[b(a-X) + \int \frac{(b-S)^2 dz}{s}\right]; \tag{34}$$

for shortness, let us write $\frac{(b-S)^2}{s} = \Phi$, hence, the living force is

$$V^2[b(a-X) + \int \Phi dz], \tag{35}$$

whose differential originated from both variabilities will be

$$2VdV[b(a-X) + \int \Phi dz] - bV^2 dX + V^2 \Phi dz + V^2 dX \int dz\left(\frac{d\Phi}{dX}\right), \tag{36}$$

on the other hand, there will be

$$\left(\frac{d\Phi}{dX}\right) = -2\frac{(b-S)}{s}\left(\frac{dS}{dX}\right) - \frac{(b-S)^2}{s^2}\left(\frac{ds}{dX}\right), \tag{37}$$

---

[20] TN: it should be $XP$, instead of $ZV$.

and since

$$S = \int dz \left(\frac{ds}{dX}\right) \quad \text{there will be also} \quad \left(\frac{dS}{dX}\right) = \int dz \left(\frac{d^2s}{dX^2}\right).$$

Therefore, once these values are substituted, the increment of the living force, after putting $dz = \frac{(b-S)dX}{s}$, will be

$$2VdV[b(a-X) + \int \Phi dz] - bV^2 dX + V^2 \frac{(b-S)^3}{s^2} dX - 2V^2 dX \int dz \frac{(b-S)}{s} \int dz \left(\frac{d^2s}{dX^2}\right) - V^2 dX \int \frac{(b-S)^2}{s^2} \left(\frac{ds}{dX}\right). \tag{38}$$

Or, since the function $\Phi$ is expected to be known, this increment can be expressed succinctly as

$$2VdV[b(a-X) + \int \Phi dz] - bV^2 dX + \frac{V^2(b-S)^3}{s^2} dX + V^2 dX \int dz \left(\frac{d\Phi}{dX}\right). \tag{39}$$

38. The increment of the living force is due, first, to the pressure $P$, applied to the basis $= b$, giving the force $= bP$, which acting through the infinitesimal distance $dX$, it should result in the formula $4gbPdX$, as we saw earlier. Because, simultaneously, the same mass of fluid is pushed backwards by the pressure $p$ applied to the basis $= s$, the resulting force will be $ps$, and acting through the infinitesimal distance $dz = \frac{(b-S)dX}{s}$, gives $4gp(b-S)dX$. It is now necessary to subtract this expression from the previous one, to obtain the formula for the increment of the living force, which, in fact, will be $4gdX[b(P-p) + pS]$, whence resulting in the following equation

$$4g[b(P-p) + pS] = \frac{2VdV}{dX}[b(a-X) + \int \Phi dz] - bV^2 + V^2 \frac{(b-S)^3}{s^2} + V^2 \int dz \left(\frac{d\Phi}{dX}\right), \tag{40}$$

being necessary to remember that

$$S = \int dz \left(\frac{ds}{dX}\right) \quad \text{and} \quad \Phi = \frac{(b-S)^2}{s}.$$

39. However, once this equation was found, it should be pointed out that it is necessary to furnish a relation between the pressure $p$ and the amplitude $s$, which we assumed earlier to be given by the formula $p = \frac{cs}{\Sigma-s}$ or else $p = c \, log \frac{\Sigma}{\Sigma-s}$, where $\Sigma$ denotes the maximum amplitude to which the tube can expand in $Z$, such that $\Sigma$ is a function of $z$ itself, and independent of the variable $X$. However, if in our equation we write such expression in the place of $p$, then our equation will depend on two variables $X$ and $z$, and, first, the quantities $P$ and $V$ will be functions of $X$ only, whereas the letters $s$, $S$ and $\Phi$ indicate that these are simultaneous functions of both variables $X$ and $z$, being only the letter $\Sigma$ a function of $z$ itself.

40. Therefore, the solution of the equation found by us, involves one more determination. Firstly, it is necessary to deduce the particular relation between $X$ and $V$, which is held constant, so that the other variable $z$, and the other dependent quantities are varied. Thereafter, once this relation is established, then $X$ is substituted for $V$, and from the same equation, the nature of the function $s$ can be drawn, so that $s$ can be determined from the variable $X$.

41. However, for the relation between $V$ and $X$ to be established, it is necessary that the pressure of the fluid in a certain location of the tube, such as $MN$, can be considered to be known, such as at the extremities of the arteries, where they terminate in glandules or veins, where it is possible to guess the pressure. Take then, the section $MN$ at the extremity of our tube, and let us write the whole interval $CM = f$, where the pressure remains always constant $= k$, henceforth, the amplitude at this location will be also constant and $= h$. In this case, once we put $z = f$, $s = h$, and $p = k$ in our equation, the velocity $V$ can be determined from the functions that depend on $X$ itself , whose [value], once substituted again in the general equation, will give an equation, which will be suited to find the nature of the function $s$, and certainly, what will be the function of the two variables $X$ and $z$.

42. But, because the possibility of a straightforward solution to be obtained is null, since this investigation is considered to transcend the human forces, so that we are forced to put an end to this work. Nonetheless, meanwhile, perhaps, it will be convenient to consider the circumstance, when a new action of the heart and the same condition of the blood through all the arteries take place, whose condition will be again initiated when the action of our piston has ceased, which is equivalent to put[21] $X = h$, and all the quantities in our equation are straightforwardly restored again to the same values, which they had at the beginning when $X = 0$, and thus all the functions that appear can thus be compared, for $X = 0$ and for $X = a$, when the very same distinct values are attained.

43. Thus, in solving for the motion of the blood, we encounter insurmountable difficulties, which hindered the thorough and precise investigation of all the works of the Creator; wherein we ought to much more constantly admire and to venerate the supreme wisdom and omnipotence, since, truly, not even the greatest human ingenuity avails to understand and to explain the true structure of even the slightest micro-organism.

____________________________

Final Comments

What was really Euler' contribution to the study of blood flow through arteries? From what we saw, it seems that his final statements speak for themselves; in summary they reveal that, despite all the efforts, he was unable to advance into his initial proposals. Nonetheless, he shows a great scientific sensibility in bringing up the concern of a theoretical approach to the problem, and to show all the difficulties to advance the modeling[22].

As far as the solution for rigid tubes is concerned, his approach is impeccable. By invoking the continuity equation and the momentum equation, and with great analytical skill, he was able to reduce the problem to the integration of a single differential equation for a single unknown, for the flow of liquids through rigid tubes with variable cross sections. From this, he gave closed form expressions for the velocity and pressure for tubes of uniform cross sections. He then, rederived his general equation by applying the work kinetic energy theorem, which Euler's called the method of the living forces (akin to kinetic energy).

Finally, as earlier pointed out[23], Euler will return to this matter again, in the memoir of 1754 (E206), where he considers the problem of raising water to an elevated reservoir with a piston pump, where he derives an expression for the pressure in any location along the pipe, which, with the exception of the gravity terms and different notation, is exactly in the same form of his general equation as presented earlier.

---

[21] TN: it should be $X = a$, instead of $X = h$.
[22] For a summary of the evolution of bloodstream modeling since Euler's first attempt see *op. cit. supra* Note 1.
[23] *Op. cit. supra* Note 2.